\documentstyle[preprint,revtex]{aps}
\begin{document}
\begin{title}
Impurity systems fluctuating between two magnetic configurations:\\
candidates for non Fermi-liquid behavior
\end{title}
\author{A.A. Aligia and M. Bali\~{n}a}
\begin{instit}
Comisi\'{o}n Nacional de Energ\'{\i}a At\'{o}mica\\
Centro At\'{o}mico Bariloche and Instituto Balseiro\\
8400 S.C. de Bariloche, R\'{\i}o Negro, Argentina\\
\end{instit}
\begin{abstract}
The appropriate generalization of the isotropic impurity Anderson
model for valence fluctuations between two magnetic multiplets
$l^n$ and $l^{n+1}$ is solved in the strong-coupling limit of
Wilson's renormalization group for $l\leq$ 3.  Except in the
extreme case of $j-j$ coupling, the ground state is degenerate, the
impurity magnetic moment is very small and is overscreened by the
conduction electrons.   The strong-coupling fixed point turns out
to be unstable.  Thus, at low enough temperatures the physics of
the systems should be governed by an intermediate coupling fixed
point and non Fermi-liquid behavior is expected.
\end{abstract}
\pagebreak
\par The Anderson model for dilute magnetic alloys and its integer
valent limit, the Kondo problem, have played a decisive role in the
understanding of intermediate-valence and heavy-fermion systems.
The scaling behavior of these models has been established
\cite{and70,wil75,kris80,kris}.  The low temperature properties are
described by the strong-coupling fixed point, characterized by a
non-magnetic ground state and Fermi-liquid behavior.  These models
and their extension to an arbitrary degeneracy of the magnetic
configuration have been solved exactly using the Bethe ansatz
\cite{schlott}.  The predicted thermodynamic properties are in
general in good agreement with experiments in dilute systems
fluctuating between one magnetic and one non-magnetic configuration
\cite{schlott}.
\par The properties of systems in which both accessible
configurations are magnetic are markedly peculiar.  For example
TmSe is the only intermediate valence compound which orders
magnetically with a local magnetic moment 1.5$\mu_B$ \cite{bje77},
much smaller than those of the 4f$^{12}$ and 4f$^{13}$
configurations.  The compounds UPt$_3$ \cite{ae88} and UBe$_{13}$
\cite{klei90} also order magnetically with a magnetic moment of the
order of 10$^{-2} \mu_B$.  The magnetic susceptibility of dilute Tm
systems points towards a magnetic ground state of Tm impurities
\cite{ber79}.  The magnetic susceptibility of the
Y$_{0.8}$U$_{0.2}$Pd$_3$ also diverges at temperature $T
\rightarrow 0$ and a finite entropy at $T=0$, $S(0)$, is suggested by
specific heat measurements \cite{sea91,and91}.  D.L. Cox stated
that the different behavior of U and Ce systems might be caused by
different underlying symmetries, and constructed an Anderson-like
model starting from crystal-field split levels \cite{cox87}.  In
the limit in which only the $\Gamma_3$ non-magnetic doublet of the
4f$^2$ configuration and the $\Gamma_7$ states of the excited
4f$^1$ configuration are retained and also the conduction $j=7/2$
partial-wave states are omitted, his model takes the form of the
spin 1/2 two-channel Kondo problem \cite{noz80}.  The
low-temperature properties of the overscreened multi-channel Kondo
problem are characterized by an intermediate-coupling fixed point
\cite{noz80,crag80} with a non-Fermi-liquid behavior
\cite{crag80,lud91} which might be related to the marginal Fermi
liquid phenomenology of the high $T_C$ systems \cite{var89}.  The spin
1/2 two-channel Kondo problem has been solved exactly \cite{lud91}
and is consistent with several properties observed in
Y$_{0.8}$U$_{0.2}$Pd$_3$, such as the logarithmic divergence of the
specific heat at low temperatures with a residual entropy $S(0)$
$\sim$ (1/2) ln2 \cite{sea91}.  However, other properties, such as
the magnetic field dependence of the specific heat, are
inconsistent with the model \cite{and91}.  The same happens with
some data of UBe$_{13}$ \cite{cox87}.
\par The study of microscopic models with non-Fermi-liquid
behavior, particularly if they are relevant to real systems, is of
great interest at present.  In this letter we show that neglecting
crystal fields, the low-temperature properties of systems
fluctuating between two p, d or f configurations are characterized
by an intermediate-coupling fixed point, except in the unrealistic
extreme cases in which one of both hybridization channels $V_j(j=l
\pm 1/2$, see below) vanishes.  Unfortunately, only one $V_j$ was
retained in most previous studies of valence fluctuations between
two magnetic configurations \cite{al86,lust82,read86,nunes87,ya85}.
 In particular $V_{l+1/2} V_{l -1/2} = 0$, if $j-j$ coupling (instead
of the more realistic $LS$ coupling) is assumed.  Also, as we have
shown elsewhere \cite{bal90}, there are not enough conduction
electron degrees of freedom in the variational wave functions
considered in Ref. \cite{ya85}, to capture the essential physics of
Tm impurities.
\par The generalized Anderson model for an impurity ion in an
isotropic medium can be written in the form:
\begin{equation}
H = H_{band} + H_{mix} + H_{ion}
\end{equation}
The first term describes a band of extended states and we need to
retain only states with the same orbital angular momentum as that
of the impurity shell $l$
\begin{equation}
H_{band} = \sum_{k \mu \sigma} \epsilon_k c^{\dag}_{k \mu \sigma}c_{k
\mu \sigma} = \sum_{kjm} \epsilon_k c^{\dag}_{kjm} c_{kjm}
\end{equation}
where $\mu, \sigma$ are the orbital angular momentum projection and
spin respectively.  The last expression is an alternative
description in terms of the total angular momentum $j=l \pm 1/2$
and its projection $m$.  Similarly:
\begin{equation}
H_{mix} = V \sum_{k \mu \sigma} (f^{\dag}_{\mu \sigma} c_{k \sigma} +
h.c) = V \sum_{kjm} (f^{\dag}_{jm} c_{kjm} + h.c.)
\end{equation}
where the $f$ operators refer to impurity electrons.  $H_{ion}$
describes the impurity shell and contains all the correlations of
the problem.  We retain only the ground state multiplets of two
neighboring $l^n$ and $l^{n+1}$ configurations:
\begin{equation}
H_{ion} = E \sum_{M_0} \mid J_0M_0><J_0M_0\mid + (E+ \Delta)
\sum_{M_1} \mid J_1M_1><J_1M_1 \mid
\end{equation}
where $J_i$, $M_i$ are the total angular momentum and projection of the
$l^{n+i}$ configuration.  We find more convenient to express the $f$
operators in terms of the Hubbard operators $\mid J_0M_0><J_1M_1
\mid$.  When this is done $H_{mix}$ takes the form:
\begin{equation}
H_{mix} = \sum_{kjmM_0M_1} V_j <J_0jM_o m \mid J_1M_1>
(c^{\dag}_{kjm} \mid J_0M_0><J_1M_1 \mid + h.c.)
\end{equation}
where the $<J_0jM_0m \mid J_1M_1>$ are Clebsch-Gordan
coefficients.  The $V_j/V$ are not free parameters, but are
determined by the particular form of the highly correlated ionic
states.  In the simplest case of $j-j$ coupling, it is easy to verify
that for $0 \leq n \leq 2l - 1$:
\begin{equation}
 V_{l + 1/2} = 0; V_{l-1/2} = V <
J_0 (l-1/2) J_0 (l + 1/2 - n) \mid J_1 J_1 >^{-1}
\end{equation}
while for $2l \leq n
\leq 4l+2$:
\begin{equation}
V_{l-1/2}=0; V_{l+1/2} = V < J_0 (l+ 1/2) J_0 (3l+1/2-n) \mid J_1 J_1>^{-1}
\end{equation}
Instead, using $LS$ coupling and the Hund rules, it can be shown that
\cite{bal}:
\begin{equation}
\frac {V_j}{V} = \frac{\sum_{\mu \sigma
L^z_iS^z_i}<L_0S_0L^z_0S^z_0 \mid J_0J_0><L_1S_1L^z_1S^z_1 \mid
J_1J_1><l {1 \over 2} \mu \sigma \mid j(J_1 - J_0)><L_0 l L^z_0 \mu
\mid L_1 l L_1^z><S_0 {1 \over 2} S^z_0 \sigma \mid S_1S^z_1>}
{<J_0j J_0 (J_1-J_0) \mid J_1J_1><L_0 lL_0 (L_1 -L_0) \mid L_1
L_1><S_0 {1 \over 2} S_0 (S_1 - S_0) \mid S_1 S_1>}
\end{equation}
The result of Eq.(8) for the cases in which both $l^n$ and
$l^{n+1}$ configurations are magnetic is given in Table I.
\par The integrability of the model using the Bethe ansatz has been
studied by Aligia et al. \cite{al86}.  When both configurations are
magnetic, only the cases with $l=1$ (see Table I) are integrable if
in addition $V_{3/2} = 0$ is assumed.  They were solved by the same
authors \cite{al86} and by Schlottmann \cite{schlott}.  In the
integer-valent limits $\mid \Delta \mid >>V$, these cases are
equivalent to underscreened Kondo problems \cite{al86},
characterized by the same strong coupling fixed point at low
temperatures as the original model \cite{all84}.  The ground state
has total angular momentum $min (J_0,J_1)$.  It has been shown
recently that in the underscreened Kondo problem, Fermi-liquid
excitations coexist with the magnetic degrees of freedom and that
they are decoupled at low frequency \cite{gan92}.
\par Unfortunately, for the most realistic cases for valence
fluctuations between two magnetic configurations the model is not
integrable with the Bethe ansatz and also alternative, accurate
enough methods used for $l=0$, have serious practical limitations.
The nature of the correlations of $H_{ion}$ would require several
different Hubbard-Stratonovich transformations for the
implementation of a Monte Carlo algorithm \cite{hir86}, while in
Wilson's renormalization group \cite{wil75,kris80,kris}, the size
of the relevant Hilbert space is enlarged by a factor $2^{4l+2}$ in
each interation, severely affecting the numerical accuracy for $l
\neq 0$.  A similar, although not so severe restriction was
encountered in the renormalization group treatment of the spin 1/2
two-channel Kondo problem \cite{crag80}.  However, formally,
Wilson's renormalization group can be implemented in the same way
as for $l=0$ \cite{wil75,kris80,kris}.  The only difference is that
the logarithmic discretized band has orbital degeneracy.  We call
$c_{n \mu \sigma}$ the states and destruction operators of the
different onion-like spherical Wannier functions ($f_{n \mu}$ in
the notation of Ref. 3).  It is clear that if one starts from a
small value of $V$, during the first few iterations the system is
near the free-orbital fixed point and the effective value of $V$
increases.  If in the subsequent renormalization group flow the
system does not reach a fixed point at an intermediate value of $V$,
then, the low temperature behavior is described by the
strong-coupling fixed point $V \rightarrow \infty$, as in the case
of $l=0$ \cite{wil75,kris80,kris} or a particular model for valence
fluctuations between two magnetic configurations \cite{all84}.
However, if the strong-coupling fixed point is unstable, there must
exist at least one intermediate-coupling fixed point, as in the
spin 1/2 two-channel Kondo model \cite{noz80,crag80}.  In this
case, there are no asymptotic scattering states \cite{crag80} and a
non-Fermi-liquid behavior is expected.
\par Thus, the study of the strong-coupling fixed point might
provide a clue about what kind of physics is expected in the  model
at low temperatures.  The properties of this fixed-point are
determined by the local Hamiltonian $H_0$ in which only the
innermost Wannier functions $c_{0 \mu \sigma}$ are considered and
all energies are neglected compared to $V$ \cite{wil75,kris80,kris}.
This one-site problem is far from trivial for realistic systems.
The size of the Hilbert space of $H_0$ is ($J_0 +J_1+1) 2^{4l+3}$.
We have found the ground state of $H_0$ in  each subspace of a
given number of particles $n+n_c$ and total angular momentum
projection $J^z_t$, using the Lanczos method and a modified version
of it \cite{dag88}.  The quantum numbers of the ground state are
given in Table I.  They are independent of $\Delta$ and $V_{l \pm
1/2}$ as long as
both $V_j \neq 0$.  The cases with the same $l$ and $J_0$ and $J_1$
interchanged are related by a special electron-hole transformation
\cite{al86,bal90}.  In general, the ground state is highly
degenerate.  There are states differing not only in $J_t^z$, but
also in $J_t$ and $n_c$ with the ground state energy. Exceptions
are the cases with $l=3$ and $1 \leq n \leq 4$ where only one value
of $n_c$ and $J_t$ characterizes the ground state.  However, due to
numerical uncertainties, we cannot completely rule out the
possibility that states with other values of $J_t$ are degenerate
with the ground state in these cases.
\par The expectation value at the impurity magnetization $J^z_{imp}
= \sum_{i,Mi} M_i \mid J_i M_i>$ is always negative for any state
of the ground state with maximum $J^z_t$.  Some specific values are
given in Ref. \cite{bal91}.  Thus, the impurity magnetization is
overscreened.  The absolute value of $<J^z_{imp}>$ decreases when the
magnitude of the $V_j$ are changed from the ideal $LS$-coupling
values to those corresponding to intermediate coupling, and
vanishes in the $j-j$ coupling limit.  In this limit the degeneracy
of the ground state becomes irrelevant, since the states $c_{ojm}$
with $j$ such that $V_j =0$ become decoupled.  If one is almost in
the $j-j$ coupling limit, say $V_{j1} >> V_{j2}$, where $V_{j2}$
would vanish according to Eqs. (6), (7), there will exist a regime
at intermediate temperatures characterized by a fixed point in
which $V_{j1} = \infty$, $V_{j2} =0$.  Our results for this fixed
point agree completely with those obtained by Lustfeld for $l=3$ and
several values of $n$ in the integer valent limit of the present
model \cite{lust82}.  In particular the ground state is
non-magnetic, in agreement also with Refs. \cite{read86,nunes87}.
However, this fixed point is unstable if $V_{j2}>0$, and a finite
though small impurity magnetic moment as found in $T_m$
\cite{bje77} and U \cite{ae88,klei90} compounds starts to develop at lower
temperatures.
\par For $l$=1 and $V_{3/2}=0$ we find a ground state with $J_t = min
(J_0,J_1)$ in agreement with Bethe ansatz results
\cite{al86,schlott} and renormalization-group calculations \cite{all84}.
\par In the following we discuss the stability of the
strong-coupling fixed point.  In the immediate neighborhood of this
fixed point, eliminating the potential scattering by means of the
mapping explained in Ref. \cite{kris80} Appendix A, section 4, the
Hamiltonian has a form like:
\begin{equation}
H \stackrel {\sim}{=} H_0 (V) + H'_{band} + \sum_{\mu \sigma} (c^{\dag}_{0\mu
\sigma} c_{1 \mu \sigma} + h.c.)
\end{equation}
where $H_0(V)$ is the problem solved above with a very large but
finite $V$ and $H'_{band}$ represents the band without the $c_{0\mu
\sigma}$ states, discussed in Section III.B.3 of Ref. \cite{kris80}.
 In most of the present cases in which different values of $n_c$
are present in the ground state, the operators $c^+_{0 \mu \sigma}$
and $c_{0 \mu \sigma}$ connect different states with the ground
state energy.  In each renormalization group iteration there is, by
definition, an overall factor $\Lambda^{1/2}$ where $\Lambda >1$ is
the parameter of the logarithmic discretization of the band (Eq.~(2.22) of
Ref.~2).  Also the operator $c_{1 \mu \sigma}$ introduces a factor
$\Lambda^{1/4}$ in each iteration (Eq. (4.23) of Ref.~2).  Thus,
the last term of Eq.~(9) is accompanied by a factor $\Lambda^{1/4}$
and is a {\it relevant} operator, indicating that the
strong-coupling fixed point is unstable.
\par The cases in which only one $n_c$ is present in the ground
state ($l$=3, $n$=1 or 2 in Table I) are more delicate, but are
completely analogous to the corresponding case of the spin-1/2
two-channel Kondo model \cite{noz80,crag80}.  In these cases,
the last term of Eq.~(9) affects the eigenvalues of $H$ in second
order perturbation theory, leading to an effective interaction
$H_{eff}$ between the ground state of $H_0$ and the $c_{1 \mu
\sigma}$ states.  Following a similar reasoning to that which led
to the effective exchange Hamiltonians in the integer valent limits
of the present model \cite{al86,lust82}, one can show that
$H_{eff}$ has the form of an exchange Hamiltonian between the total
angular momentum $J_t$ of the ground state of $H_0$ and the
$c_{1jm}$ states.  Also since the relevant denominators in the
perturbative treatment should increase like $V$ as $\Lambda^{1/2}$ on
each iteration, $H_{eff}$ is a marginal operator.  Following
Ref.~2, we expect that the strong-coupling fixed point is stable if
the coupling constant $J_{eff}$ of $H_{eff}$ is negative
(ferromagnetic) and unstable otherwise.  A simple argument due to
Nozi\`{e}res and Blandin \cite{noz80} shows that if the impurity
magnetic moment is over-(under) screened in $H_0$, $J_{eff} >0$
($J_{eff} < 0$).  Thus, the strong-coupling fixed point is also unstable.
\par In conclusion, valence fluctuations between two magnetic $l^n$
and $l^{n+1}$ configurations, except in limiting unrealistic cases
in which $V_{l+1/2} V_{l-1/2} = 0$ (e.g. extreme $j-j$ coupling) are
epected to be characterized by an intermediate coupling fixed point
at low temperatures, in absence of crystal field.
Although the cases with $l=1$ might look academic,
they are the simplest ones to be studied by alternative methods,
which we hope will follow the present study.  The unrealistic case
$l=0$ has already provided a great deal of understanding of the
physics of systems fluctuating between one magnetic and one
non-magnetic configurations.
\par We would like to thank Andrei E. Ruckenstein for many
enlightening discussions, and L.M. Falicov for a critical reading
of the manuscript.  One of us (MB) is supported by the Consejo
Nacional de Investigaciones Cient\'{i}ficas y T\'{e}cnicas
(CONICET).  AAA is partially supported by the CONICET.

\begin{table}
Parameters and ground-state quantum numbers in the strong coupling
limit, for valence fluctuations between any two magnetic Hund-rules
ground-state multiplets of neighboring $l^n$ and $l^{n+1}$
configurations with $l \leq 3$. $J_i$ is the total angular momentum
of the $l^{n+i}$ configuration and $V_j$ is the hybridization
energy for conduction electrons with total angular momentum $j$,
using $LS$ coupling.  The numbers in brackets are those $V_j$ which
vanish for $j-j$ coupling.  The total number of particles and the
total angular momentum of the local Hamiltonian $H_0$, are denoted
by $n+n_c$ and $J_t$ respectively.  Some cases of marginal interest
with $l=3$ were not solved.
\begin{tabular}{c|c|cc|cc|cl}
$l$ & $n$ & $J_0$ & $J_1$ & $V_{l-1/2}/V$ & $V_{l+1/2}/V$ & $n_c$ &
$J_t$ \\ \hline
1 & 3 & 3/2 & 2 & (0.8165) & 0.8165 & 2-4 & 0,1/2 \\
1 & 4 & 2 & 3/2 & (0.6455) & 1.2910 & 3-5 & 0,1/2  \\ \hline
2 & 1 & 3/2 & 2 & 1.1593 & (0.5059) & 3-7 & 0,3/2,2 \\
2 & 2 & 2 & 3/2 & 1.0733 & (0.7026) & 4-8 & 0,3/2,2 \\
2 & 5 & 5/2 & 4 & (0.8944) & 0.6324 & 5-9 & 0,3/2,2 \\
2 & 6 & 4 & 9/2 & (0.9083) & 0.9710 & 4-8 & 0,3/2,2 \\
2 & 7 & 9/2 & 4 & (0.7817) & 1.2536 & 3-7 & 0,3/2,2 \\
2 & 8 & 4 & 5/2 & (0.5477) & 1.5492 & 2-6 & 0,3/2,2 \\ \hline
3 & 1 & 5/2 & 4 & 1.2234 & (0.4666) & 8 & 15/2 \\
3 & 2 & 4 & 9/2 & 1.1848 & (0.6971) & 8 & 8 \\
3 & 3 & 9/2 & 4 & 1.0853 & (0.8514) & 7 & 8 \\
3 & 4 & 4 & 5/2 & 0.9587 & (0.9141) & 7 & 15/2 \\
3 & 7 & 7/2 & 6 & (0.9258) & 0.5345 & 7-13 & 0,3/2,2,5/2,4,9/2 \\
3 & 8 & 6 & 15/2 & (1.0059) & 0.8138 & & \\
3 & 9 & 15/2 & 8 & (0.9647) & 1.0362 & & \\
3 & 10 & 8 & 15/2 & (0.8612) & 1.2334 & & \\
3 & 11 & 15/2 & 6 & (0.7058) & 1.4275 & & \\
3 & 12 & 6 & 7/2 & (0.4818) & 1.6690 & 2-8 & 0,3/2,2,5/2,4,9/2 \\
\end{tabular}
\end{table}

\end{document}